\documentclass[epj]{svjour}
\usepackage{epsfig} 
\usepackage{cite} 
\begin{document}  
\title{Collisional deexcitation of  exotic hydrogen atoms in highly excited states. 
I. Cross-sections
}
\titlerunning{Collisional deexcitation of  exotic hydrogen atoms. I. Cross-sections}
\author{ T.S. Jensen\inst{1,2} \and V.E. Markushin\inst{1}}
\institute{
Paul Scherrer Institute, CH-5232 Villigen PSI, Switzerland
\and 
Institut f{\"u}r Theoretische Physik der Universit{\"a}t Z{\"u}rich,
Winterthurerstrasse 190, CH-8057 Z{\"u}rich, Switzerland 
}
\date{Received: date / Revised version: date}
\abstract{
The deexcitation of exotic hydrogen atoms in highly excited states 
in collisions with hydrogen molecules has been studied using 
the classical-trajectory Monte Carlo method.  The Coulomb transitions 
with large change of principal quantum number $n$ have been found to be the 
dominant collisional deexcitation mechanism at high $n$.  The molecular 
structure of the hydrogen target is shown to be essential for the dominance 
of transitions with large $\Delta n$. The external Auger effect has been
studied in the eikonal approximation. The resulting partial wave cross-sections 
are consistent with unitarity and provide a more reliable  input for cascade calculations than
the previously used Born approximation.
\PACS{
      {34.50.-s}{Scattering of atoms and molecules}   \and
      {36.10.-k}{Exotic atoms and molecules (containing mesons, muons, 
      and other unusual particles)}
     } % end of PACS codes
}

\maketitle

\newcommand{\be}{\begin{equation}}
\newcommand{\ee}{\end{equation}}
\newcommand{\non}{\nonumber}
\newcommand{\dd}{\mbox{\rm d}}
\newcommand{\mup}{\mbox{$\mu^-p$}}
\newcommand{\mud}{\mbox{$\mu^-d$}}
\newcommand{\pip}{\mbox{$\pi^-p$}}
\newcommand{\pbp}{\mbox{$\bar{p}p$}}
\newcommand{\Kp}{\mbox{$K^-p$}}
\newcommand{\eV}{\mbox{\rm eV}}
\newcommand{\xp}{\mbox{$x^-p$}}
\newcommand{\muxp}{\mbox{$\mu_{xp}$}}
\newcommand{\Ha}{\mbox{${\mathrm{H}}$}}   
\newcommand{\Vxp}[1]{V_{x^-p}(#1)}
\newcommand{\VxpH}[2]{V_{x^-p-H}(#1,#2)}
\newcommand{\VeH}[1]{\left(\frac{1}{#1}+1\right)e^{-2#1}} 
\newcommand{\VHH}[1]{V_\mathrm{HH}(#1)}
\newcommand{\rb}{\mathbf{r}}
\newcommand{\Rb}{\mathbf{R}}
\newcommand{\rhomax}{\rho_{\mathrm{max}}}

%%%%%%%%%%%%%%%%%%%%%%%%%%%%
%\begin{flushright}
%jm2001hx\_h \quad December 30, 2001
%\end{flushright}
%%%%%%%%%%%%%%%%%%%%%%%%%%%%

%%%%%%%%%%%%%%%%%%%%%%%%%%%%%%%%%%%%%%%%%%%%%%%%%%%%%%%%%%%%%%%%%%%%%%%%%%%
\section{Introduction}
\label{sect:intro}
 
  Exotic hydrogen atoms $x^-p$ ($x^-=\mu^-, \pi^-, K^-, \bar{p}$) 
are formed in highly excited states with the principal quantum number 
$n\sim\sqrt{\muxp/m_e}$ where \muxp\ is the reduced mass of the exotic atom 
\cite{leon62,cohen99}.  
For a long time the initial stage of the atomic cascade remained 
poorly understood despite a substantial progress in  
theoretical and experimental studies (see \cite{borie80,markushin94,markushin99} and references 
therein).  In particular, the dominant collisional deexcitation mechanism  
was unclear for 40 years since the so-called chemical deexcitation was 
introduced in \cite{leon62} 
as a phenomenological solution to the problem of the 
cascade time at high $n$ (the external Auger effect 
alone would give much longer cascade times). 
A shortage of experimental data related to the initial stage of the 
atomic cascade hindered theoretical studies of this problem. 
The experimental situation, however, changed recently 
as more data on the atomic cascades in exotic hydrogen atoms 
at low density became available.   
The cascade time of antiprotonic hydrogen measured by the OBELIX collaboration 
\cite{obelix00} in the density range $3-150\;$mbar was found to be 
significantly shorter than the prediction of the conventional cascade 
model \cite{reifenrother89}. 
The new experimental results on the atomic cascade in muonic hydrogen 
from the PSI experiment \cite{kottmann99} provided detailed information not 
only on the cascade time, but also on the energy distribution at the 
end of the cascade, which at low density is actually preserved from the initial 
stage after the fast radiative deexcitation takes over the collisional processes. 
 
The goal of this paper is to investigate the collisional deexcitation 
mechanisms for highly excited exotic atoms. 
In particular, we are interested in the role of the Coulomb acceleration  
in highly excited states and in the competition between the acceleration 
and slowing down in quasi-elastic collisions.  
Both molecular and atomic hydrogen targets were used in our calculations 
in order to investigate the role of molecular effects.  

 The paper is organized as 
follows.  The classical-trajectory Monte  Carlo method is described in 
Section~\ref{sect:ctmc}.  The results of calculations of Coulomb, Stark, and 
transport cross-sections for the $\mu^-p$ and $\bar{p}p$ atoms are presented 
in Section~\ref{sect:res}.  The Auger deexcitation is discussed in 
Section~\ref{sect:auger}. 
The conclusions are summarized in Section~\ref{sect:conc}.   

Unless otherwise stated, atomic units ($\hbar=e=m_e=1$) are used throughout 
this paper. The unit of cross-section is $a_0^2=2.8\cdot10^{-17}\;{\rm cm}^2 $, 
where $a_0 = \frac{\hbar^2}{m_e e^2}$ is the electron Bohr radius.

%%%%%%%%%%%%%%%%%%%%%%%%%%%%%%%%%%%%%%%%%%%%%%%%%%%%%%%%%%%%%%%%%%%%%%%%%%%
\section{Classical-trajectory  Monte Carlo calculation}
\label{sect:ctmc}

%%%%%%%%%%%%%%%%%%%%%%%%%%%%%%%%%%%%%%%%%%%%%%%%%%%%%%%%%%%%%%%%%%%%%%%%%%%
\subsection{Effective potential}

In the beginning of the atomic cascade, where many $nlm$-states
are involved in the collisions, classical mechanics is expected 
to be a good approximation.  To study 
the scattering of exotic hydrogen atoms from hydrogen atoms or molecules 
% the collisional processes at the initial stage of the atomic cascade 
we use a classical-trajectory Monte Carlo model. 
   The following degrees of freedom are included in the model: the constituents 
of the exotic atom ($x^-=\mu^-,\pi^-,K^-,\bar{p}$ and the proton) and 
the hydrogen atoms are treated as classical particles. The electrons 
are assumed to have fixed charge distributions corresponding to the 
$1s$ atomic state around the protons in the hydrogen atoms. 

\begin{figure}
\center{
\mbox{
\epsfig{file=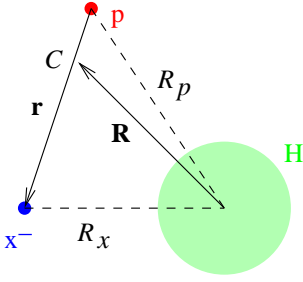, width=4.0cm}
}
}
\caption{\label{fig:coord1}
 Coordinates for the three-body subsystem. $C$ is the center of mass of the
\xp. 
}
\end{figure}

\begin{figure}
\center{
\mbox{
\epsfig{file=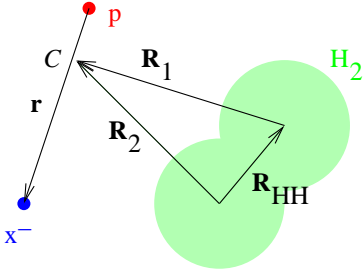, width=4.8cm}
}
}
\caption{\label{fig:coord2}
Coordinates for the four-body system.   
}
\end{figure}

   We describe the exotic atom as a classical two-body system with the 
potential 
\begin{eqnarray}
 \Vxp{\rb} & = & -\frac{1}{r} .  
 \label{Vxp}
\end{eqnarray}
The exotic atom interacts with two hydrogen atoms whose electron distributions 
are assumed to be frozen in the ground atomic state 
(see Figure~\ref{fig:coord1} for notation):
\begin{eqnarray}
 \VxpH{\rb}{\Rb} & = & \VeH{R_p} \nonumber\\
 &-& \VeH{R_x} 
\end{eqnarray}
The interaction between the hydrogen atoms is described by the Morse potential 
\begin{eqnarray}
  \VHH{\Rb_\mathrm{HH}} & = & D_e (e^{-\alpha (R_\mathrm{HH}-R_0)} - 1)^2
  \label{vmorse}
\end{eqnarray}
where $D_e=4.75~\mathrm{eV}$, $\alpha=1.03$, and $R_0=1.4$~\cite{bransden83}. 
   The effective potential for the $\xp+\Ha+\Ha$ system (see Figure~\ref{fig:coord2}) 
has the form
\begin{eqnarray}
  V & = & \Vxp{\rb} + 
          \VxpH{\rb}{\Rb_1} + 
          \VxpH{\rb}{\Rb_2}\nonumber\\
   &+&    \VHH{\Rb_\mathrm{HH}} 
      . 
\label{Veff} 
\end{eqnarray}

%%%%%%%%%%%%%%%%%%%%%%%%%%%%%%%%
%%%%%%%%%%%%%%%%%%%%%%%%%%%%%%%%
%%%%%%%%%%%%%%%%%%%%%%%%%%%%%%%%
%%%%%%%%%%%%%%%%%%%%%%%%%%%%%%%%

%%%%%%%%%%%%%%%%%%%%%%%%%%%%%%%%%%%%%%%%%%%%%%%%%%%%%%%%%%%%%%%%%%%%%%%%%%%
\subsection{Method of calculation}

   The classical equations of motion corresponding to the effective potential 
(\ref{Veff}) were solved using a fourth-order Runge-Kutta method.  
The initial conditions were defined as follows.  
Given the initial 
principal quantum number $n_i$ and 
the orbital angular momentum $l_i$ of the \xp\ atom,
  the initial classical state was generated as  
a classical Kepler orbit with the total CMS energy $E_{xp}$ and the classical angular 
momentum $l_c$: 
\begin{eqnarray}   
    E_{xp} & = & - \frac{\muxp}{2n_i^2},  \label{InitCondN}
\\
        l_c & = & l_i +\frac{1}{2}.            \label{InitCondL}
\end{eqnarray}
The orbit was oriented randomly in space, and the orbital \xp\ motion was 
set at a random time within the period.  
The hydrogen atoms in the target molecule were set at the equilibrium distance 
$R_0$, and the molecule was randomly oriented in space. 
The impact parameter $\rho$ of the \xp\ atom 
was selected with a uniform distribution in the interval 
$(0,\rho_{\mathrm{max}})$, as discussed below.
        The accuracy of the numerical calculations was controlled  
by checking the conservation of total energy and angular momentum.
Instead of requiring convergence for every individual trajectory,
we  used the global criteria that the cross-sections for the various  
processes (see below) were stable within the statistical errors
against further increase in the numerical accuracy for each collision.

\begin{figure}[htb]
\center{
\epsfig{file=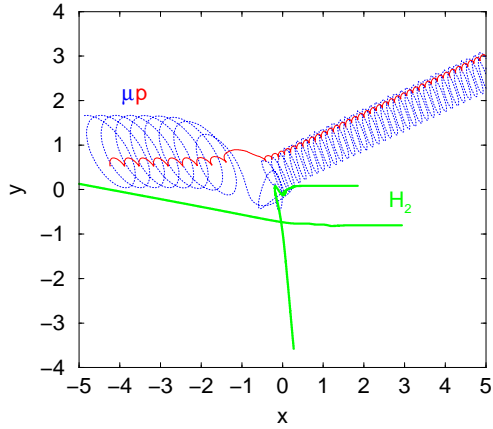, width=6.5cm}
}
\caption{
An example of a $\mup +\Ha_2 $ collision with impact parameter $\rho=a_0$
 resulting in Coulomb 
deexcitation of the \mup\ and dissociation of the $\Ha_2$. 
The exotic atom with laboratory kinetic energy  $T_i=1\;\eV$, $n_i=14$, and $l_i=13$ 
 enters from the left, the hydrogen molecule from
the right.  The trajectories are plotted in the CMS.
In the final state, the \mup\ has $n_f=10$, $l_f=7$, and $T_f=4.3\;$eV. 
}
\label{fig:collision}
\end{figure}

The final atomic state was determined when the distance between $\xp$ and 
the hydrogen atoms after the collision was larger than $10 a_0$. 
The final atomic state with the energy $E_{xp}$ and the angular 
momentum $l_c$ was identified as corresponding to the final $n_f l_f$ 
state according to the rules similar to (\ref{InitCondN},\ref{InitCondL}): 
\begin{eqnarray}   
      n_f-\frac{1}{2} < & n_c = \sqrt{2|E_{xp}|/\muxp} & \leq n_f+\frac{1}{2}
\\
      l_f < & l_c {n_f}/{n_c} & \leq l_f+1.
\label{FinalCond}
\end{eqnarray}
   In addition to the quantum numbers $n_f$ and $l_f$, the CMS scattering 
angle $\theta$ and the excitation energy of the target
$\Delta E_\mathrm{target}$ were also obtained.
For the purpose of cascade calculations, 
we are mainly interested in the reaction channels that include 
the $\xp$ atomic states: 
\begin{eqnarray}
 (\xp)_{n_i l_i} + \Ha_2 & \to & \left \{
   \begin{array}{l}
      (\xp)_{n_f l_f}+\Ha_2 \\
      (\xp)_{n_f l_f}+\Ha_2^* \\
      (\xp)_{n_f l_f}+\Ha + \Ha 
  \end{array} \right. 
\label{FinAt}
\end{eqnarray}
Other possible channels are the breakup reactions 
\begin{eqnarray}
 (\xp)_{n_il_i} + \Ha_2 & \to & \left \{
   \begin{array}{l}
      x^- + p +\Ha_2^* \\
      x^- + p +\Ha + \Ha 
  \end{array} \right. 
\label{FinBr}
\end{eqnarray}
and the formation of $(x^-\Ha)_{n_f l_f}$ ions 
\be
 (\xp)_{n_i l_i} + \Ha_2 \to 
      (x^-\Ha)_{n_f l_f} + p + \Ha. 
\label{FinMM}
\ee
%\begin{eqnarray}
% (\xp)_{n_i l_i} + \Ha_2 & \to & \left \{
%   \begin{array}{l}
%      (x^-\Ha)_{n_f l_f} + p + \Ha 
%  \end{array} \right. 
%\label{FinMM}
%\end{eqnarray}
An example of a collision that results in Coulomb deexcitation of the \mup\
and dissociation of the hydrogen molecule is shown in Figure~\ref{fig:collision}.  
%Figure~\ref{fig:collision}b shows a $\pbp+\Ha_2$ collision resulting in Coulomb
%deexcitation of the \pbp\ and excitation of the $\Ha_2$ molecule.

%%%%%%%%%%%%%%%%%%%%%%%%%%%%%%%%%%%%%%%%%%%%%%%%%%%%%%%%%%%%%%%%%%%%%%%%%%%%%
%%%%%%%%%%%%%%%%%%%%%%%%%%%%%%%%%%%%%%%%%%%%%%%%%%%%%%%%%%%%%%%%%%%%%%%%%%%%%

For a given initial $n$ of the \xp\ and laboratory kinetic energy, $T_i$,
 a set of  
the impact parameters $\rho_i,\; i=1,...,K$ with a uniform distribution in 
the interval $(0,\rhomax)$ was generated.  The value 
$\rhomax=5+2n_i^2/\muxp$ was found to be 
suitable for all cases concerned. 
The cross-sections were obtained from the computed set of trajectories 
using the following procedure.  
Let $P_i^\alpha$ be the probability that the reaction channel $\alpha$ 
corresponds to the final state in collision $i$:  
\begin{eqnarray}
 P_i^\alpha & = & \left\{
                        \begin{array}{rcl}
                          1 &,& \mbox{\rm if $\alpha$ occurred}\\
                          0 &,& \mbox{\rm otherwise}
                        \end{array} 
                  \right.
\end{eqnarray}
The cross-section for the reaction channel $\alpha$ is given by
\begin{eqnarray}
 \sigma_\alpha & = & 2\pi \rhomax \frac{1}{K} 
                    \sum_{i=1}^{K} \rho_i P_i^\alpha
\end{eqnarray}
The differential cross-sections are determined in a similar way 
by binning the corresponding intervals of variables like $z=\cos\theta$, 
where $\theta$ is the CMS scattering angle,   
and the target excitation energy $\Delta E_\mathrm{target}$.   
For instance, the differential cross-section $\dd\sigma(z)/\dd z$ 
is calculated using the relation 
\be
\frac{\dd\sigma(z)}{\dd z} \approx
  \frac{\sigma (z-\Delta z < \cos\theta < z+\Delta z)}{2\Delta z}
\ee

%%% The energy loss in elastic collisions with molecular excitation and 
%%% breakup =? 

%%%%%%%%%%%%%%%%%%%%%%%%%%%%%%%%%%%%%%%%%%%%%%%%%%%%%%%%%%%%%%%%%%%%%%%%%%%
\subsection{Special final states}

 The formation of $x^-\mathrm{H}$ ions in reaction (\ref{FinMM}) 
is an artifact of our model due to the treatment of the electrons as 
fixed charge distributions. The cross-sections for these processes turn 
out to be small, and usually the
final $n_f$ is small, so that the electron screening is not very important.
 For the purpose of cascade calculations,
one can count the $x^-\mathrm{H}$ formation as the \xp\ events  with 
the corresponding values of $n_f$, $l_f$, $\cos\theta$, and $\Delta E_\mathrm{target}$. 
%This approximation should not be used for $x^-\mathrm{H}$ in high $n_f$-states
%where electron screening is important, but it turns out that the cross-sections
%for these processes are small.  
%For the intermediate states $n_f = ?$, 
%where the $x^-\mathrm{H}$ process plays a 
%bigger role, the electron screening is less important. 
Another channel involving  $x^-\mathrm{H}$  ions 
is related to the formation of metastable molecular states like 
\be
  p(x^-\mathrm{H})_{n_fl_f}
%,\quad \mathrm{H}(x^-\mathrm{H})_{n_fl_f}+p,
%\quad\mathrm{or}
%\quad p(\mathrm{H}(x^-\mathrm{H})_{n_fl_f})
\ee
where a deeply bound $x^-\Ha$ ion forms a loosely bound state with the proton. 
These molecular states can be rather stable and often do not
dissociate  within a reasonable amount of computer time.  In our calculations 
 we consider the metastable molecular states as final states.  
We used the following criteria for the metastability: first, the collision time 
must exceed 
\be
  t_\mathrm{mol}=50/v_\mathrm{init}  \label{tmol}
\ee 
where $v_\mathrm{init}$ is the initial velocity of the \xp\ in the
laboratory system.  
With the choice of the time interval (\ref{tmol}), the colliding particles 
reach their asymptotically free final trajectories for most non-resonant    
collisions.  
%\MBF{discuss}
Second,  the $x^-$ must form a bound state with one of the hydrogen atoms and
the binding energy must not vary by more than 1\% within the time
\be
  \tau=20\frac{2\pi n_i^3}{\muxp}
\ee
which corresponds to 20 classical periods of the initial \xp\ atom.
Once metastability is reached, the event is counted as an $x^-\mathrm{H}$
event.

%%%%%%%%%%%%%%%%%%%%%%%%%%%%%%%%%%%%%%%%%%%%%%%%%%%%%%%%%%%%%%%%%%%%%%%%%%%
\section{Results}
\label{sect:res}

The classical-trajectory Monte Carlo method described in Section~\ref{sect:ctmc} 
has been used to obtain  the collisional cross-sections needed 
in  calculations of the cascades in  \mup, \pip, \Kp, and \pbp.   
The same method can also be used in a direct simulation of the atomic cascade 
without using pre-calculated cross-sections. % as explained in Section~\ref{sect:includecmc}.
For \mup\ and  \pbp\ atoms experimental data at low density 
are available for direct comparison with the 
cascade calculations \cite{jensen02next}. We will, therefore, present detailed results for 
these two cases. The initial stages  also affect the 
cascades in  \pip\ and \Kp\  because they determine the kinetic energy distribution
in the intermediate stage of the cascade where nuclear
 absorption becomes important. 
%Our main goal is to study the role of the Coulomb acceleration  
%in highly excited states and the competition between the acceleration 
%and slowing down in quasi-elastic collisions.  
%Both molecular and atomic hydrogen targets were used in our calculations 
%in order to investigate the role of molecular effects.  
  
The calculations have been done for $n_i=8-20$ for \mup, $n_i=13-35$ for \pbp\
and 9 values of the laboratory kinetic energy in the interval 
$0.05\;\eV \leq T \leq 20\;\eV$. At $T=1\;\eV$ the cross-sections have been calculated
down to $n_i=4$ for \mup\ and $n_i=8$ for \pbp.   
For each initial state $(n_i,T)$, 
1000 classical trajectories have been calculated as described above. 
The orbital quantum number $l_i$ was distributed according to 
the statistical weight.
For the purpose of illustration, a larger number of trajectories (up to 10000) have been 
calculated for some initial states in order to reduce statistical errors. 
Preliminary results have been shown in~\cite{jensen02hyp}.

We compare the results of the classical Monte Carlo (CMC)
 calculations with those of the  semiclassical
approximation. Bracci and Fiorentini \cite{bracci78} calculated the 
Coulomb cross-sections for muonic hydrogen scattering from atomic
hydrogen in a semiclassical model. Though the approach \cite{bracci78} may be unsuitable
for treating the low $n$ states, where more elaborate calculations
give much smaller values for the cross-sections~\cite{ponomarev99}, 
it can be expected to give a fair description  of the high $n$ region. 
In the case of Stark mixing we use the fixed field model~\cite{jensen02epjd} 
for comparison. In the case of molecular target, we
obtained a semiclassical estimate of the Stark cross-sections by using the 
spherical symmetric electric field corresponding to the
charge distribution of a $\Ha_2$ molecule in the ground state.

%%%%%%%%%%%%%%%%%%%%%%%%%%%%%%%%%%%%%%%%%%%%%%%%%%%%%%%%%%%%%%%%%%%%%%%%%%%
\subsection{Muonic hydrogen}
\label{sect:res:mup}
%%%%%%%%%%%%%%%%%%%%%%%%%%%%%%%%%%%%%%%%%%%%%%%%%%%%%%%%%%%%%%%%%%%%%%%%%%%
\subsubsection{Coulomb deexcitation}

\begin{figure}[htb]
\center{
\epsfig{file=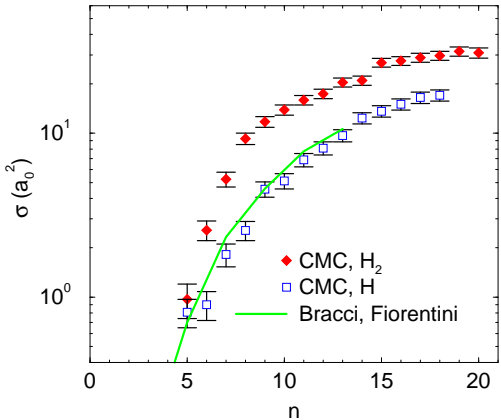, width=6.5cm}
}
\caption{The $n$ dependence of the muonic hydrogen  
Coulomb cross-sections at the laboratory energy $T=1\;$eV 
for molecular (filled diamonds) and atomic (squares) hydrogen target.   
The curve is the semiclassical result from \protect\cite{bracci78}.
}
\label{fig:cmup1}
\end{figure}

The $n$ dependences of the total cross-sections of the Coulomb deexcitation 
for collisions with molecular and atomic hydrogen  
\begin{eqnarray}
 (\xp)_{n_i l_i} + \Ha_2 & \to & \left \{
   \begin{array}{l}
      (\xp)_{n_f l_f}+\Ha_2^* \\
      (\xp)_{n_f l_f}+\Ha + \Ha 
  \end{array} \right. , 
\label{EqCoulMol}
\\
 (\xp)_{n_i l_i} + \Ha & \to & (\xp)_{n_f l_f}+\Ha  
\label{EqAtMol}
\end{eqnarray}
with $ n_f < n_i $ are shown in Figure~\ref{fig:cmup1}.  The cross-sections
increase steadily with increasing $n$ as the \mup\ becomes larger and the
energy spacing between the $n$ levels smaller. 
The cross-sections for the atomic 
target at the laboratory kinetic energy $T=1\;$eV are very close to 
the semiclassical results of Bracci and Fiorentini~\cite{bracci78}.  
The cross-section for the molecular target is larger by a factor of 
about $2 - 3$. 

%%%%%%%%%%%%%%%%%%%%%%%%%%%%%%%%%%%%%%%%%%%%%%%%%%%%%%%%%%%%%%%%%%%%%%%%%%%%%
%%%%%%%%%%%%%%%%%%%%%%%%%%%%%%%%%%%%%%%%%%%%%%%%%%%%%%%%%%%%%%%%%%%%%%%%%
%%%%%%%%%%%%%%%%%%%%%%%%%%%%%%%%%%%%%%%%%%%%%%%%%%%%%%%%%%%%%%%%%%%%%%%%%%%%%
\begin{figure}
\center{
\epsfig{file=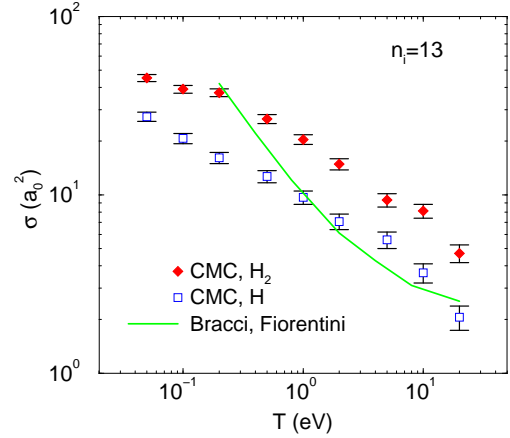, width=6.5cm}
}
\caption{
The energy dependence of the Coulomb cross-sections
for muonic hydrogen with $n_i=13$ and  
molecular (filled diamonds) and atomic (squares) hydrogen target.    
The error bars are statistical.
The curve is the semiclassical result from \protect\cite{bracci78}.
}
\label{fig:coulomb13}
\end{figure}
%%%%%%%%%%%%%%%%%%%%%%%%%%%%%%%%%%%%%%%%%%%%%%%%%%%%%%%%%%%%%%%%%%%%%%%%%%%%%
 An example of the energy dependence of the total Coulomb cross-sections
($n_f<n_i$) for  $n_i=13$ is shown in Figure~\ref{fig:coulomb13}.   
The cross-sections calculated with molecular target are approximately
twice as large as the atomic ones in the whole energy range considered. 
The CMC result for the atomic target 
is in  fair agreement with the semiclassical result \cite{bracci78} 
for energies above~1~eV. The  energy dependence of the CMC cross-sections
is approximately given by $ 1/\sqrt{T}$ corresponding to constant rates. This 
is in contrast to the $1/T$ behavior found for low energies in~\cite{bracci78}. 

\begin{figure}
\center{
\epsfig{file=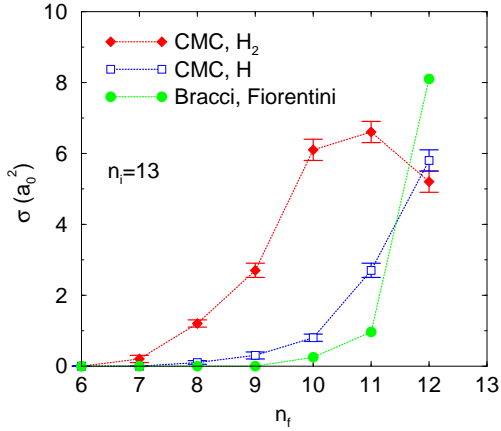, width=6.5cm}
}
\caption{
The  $n_f$ dependence of the Coulomb cross-sections 
for muonic hydrogen with $n_i=13$ and laboratory kinetic energy $T=1\;$eV 
for collisions with molecular (filled diamonds) and atomic (squares) hydrogen target.    
The semiclassical result from \cite{bracci78} is shown with filled circles.
}
\label{fig:dn13}
\end{figure}

\begin{figure}
\center{
\epsfig{file=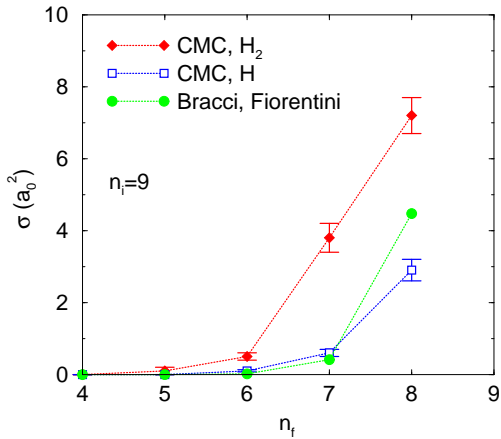, width=6.5cm}
}
\caption{
The  $n_f$ dependence of the Coulomb cross-sections 
for muonic hydrogen with $n_i=9$ and laboratory kinetic energy $T=1\;$eV 
for collisions with molecular (filled diamonds) and atomic (squares) hydrogen target.    
The semiclassical result from \cite{bracci78} is shown with filled circles.
}
\label{fig:dn9}
\end{figure}

  The distribution over final states $n_f$ is 
completely different for the molecular and the atomic targets as illustrated in  
Figure~\ref{fig:dn13} showing the $l$-average cross-sections
$\sigma_{13\to n_f}$ for $\mup$ at 1~eV. 
The calculations for atomic target predict that $\Delta n=1$ transitions
dominate the Coulomb deexcitation 
 in agreement with the semiclassical result \cite{bracci78}.
For the molecular target, the transitions with $\Delta n>1$ are strongly 
enhanced as compared to the atomic case.   
The shape of the $n_f$ distribution depends on the initial state $n_i$: 
with decreasing $n_i$ it becomes narrower and its maximum shifts towards 
smaller values of $\Delta n$. For $n_i=13$, the transitions $\Delta n=2-3$ dominate.
Figure~\ref{fig:dn9} shows the $n_f$ dependence for initial state $n_i=9$:
 the transitions with $\Delta n=1$ are most likely,
 but the $\Delta n>1$ transitions still
make up a substantial fraction of  38\% of the Coulomb cross-section as
 compared to 19\% for atomic target. 
%%%%%%%%%%%%%%%%%%%%%%%%%%%%%%%%%%%%%%%%%%%%%%%%%%%%%%%%%%%%%%%%%%%%%%%%%%%%
%%%%%%%%%%%%%%%%%%%%%%%%%%%%%%%%%%%%%%%%%%%%%%%%%%%%%%%%%%%%%%%%%%%%%%%%%
%%%%%%%%%%%%%%%%%%%%%%%%%%%%%%%%%%%%%%%%%%%%%%%%%%%%%%%%%%%%%%%%%%%%%%%%%%%%
%%%%%%%%%%%%%%%%%%%%%%%%%%%%%%%%%%%%%%%%%%%%%%%%%%%%%%%%%%%%%%%%%%%%%%%%%

%The  distribution over final $n_f$ for the Coulomb cross-sections 
%with molecular target depends on the interatomic H-H
%potential, Eq.~(\ref{vmorse}). By varying the potential depth, $D_e$, in the Morse
%potential one finds that a weaker potential shifts the maximum of the distribution
%towards smaller values of $\Delta n$. 
%A collision with a  hydrogen molecule is, therefore, qualitatively different from
%two collisions with hydrogen atoms.  

%The molecular effect on the results  has important implications 
%for the atomic cascade as discussed in Ref.~\cite{jensen02tobe}. 
%%%%%%%%%%%%%%%%%%%%%%%%%%%%%%%%%%%%%%%%%%%%%%%%%%%%%%%%%%%%%%%%%%%%%%%%%%%
\subsubsection{Stark mixing and elastic scattering}
%%%%%%%%%%%%%%%%%%%%%%%%%%%%%%%%%%%%%%%%%%%%%%%%%%%%%%%%%%%%%%%%%%%%%%%%%%%%%

\begin{figure}
\center{
\epsfig{file=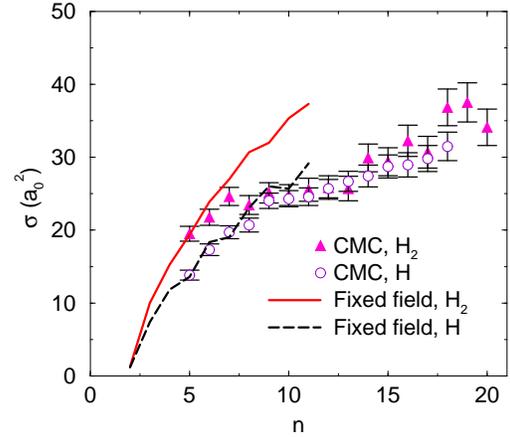, width=6.5cm}
}
\caption{
Stark cross-sections for muonic hydrogen 
 for molecular (filled triangles) and atomic (circles) hydrogen target.    
The curves show the results of the fixed field model for 
 molecular target (solid line) and atomic target (dashed line).      
The laboratory kinetic 
energy  is $T=1$~eV.
}
\label{fig:cmup2}
\end{figure}

The Stark collisions change the orbital angular momentum while preserving 
the principal quantum number: 
\begin{eqnarray}
 (\xp)_{n_i l_i} + \Ha_2 & \to & \
      (\xp)_{n_i l_f}+\Ha_2^* \; (\Ha + \Ha), 
\label{EqStMol}
\\
 (\xp)_{n_i l_i} + \Ha & \to & (\xp)_{n_i l_f}+\Ha . 
\label{EqStAt}
\end{eqnarray}
The CMC results for the $n$ dependence of the $l$-average Stark mixing 
cross-section are shown in Figure~\ref{fig:cmup2}.     
  The Stark cross-sections calculated with molecular target are less
than twice the atomic ones.  This is due to two reasons. 
First, there is a considerable molecular screening effect because the electric fields
from the two hydrogen atoms partly cancel each other. % near the center of the molecule.
Second, the 
Coulomb cross-section makes up a larger fraction of the total cross-section 
in the molecular case.  
The classical Monte Carlo results for the atomic target
are in a good agreement with the semiclassical fixed field model.  
  At low $n$, where the inelasticity due to the Coulomb deexcitation is small 
and can be neglected in the calculation of the Stark cross-sections, 
there is a good agreement between the classical Monte Carlo results 
for the molecular target and the corresponding semiclassical model. 

\begin{figure}
\center{
\epsfig{file=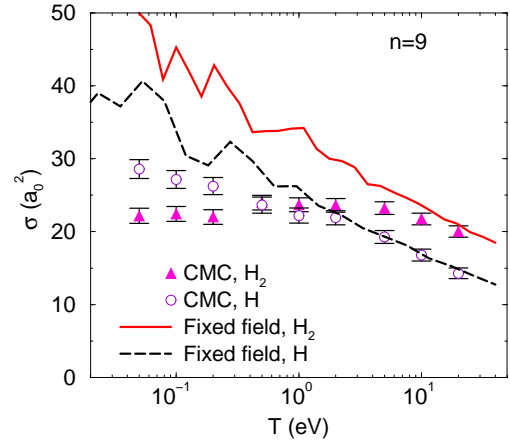, width=6.5cm}
}
\caption{
The energy dependence of the Stark cross-sections for muonic hydrogen in the state $n=9$
 for molecular (filled triangles) and atomic (circles) hydrogen target.    
The curves show the results of the fixed field model for  molecular target (solid line) 
and atomic target (dashed line).      
}
\label{fig:stark9}
\end{figure}

 Figure~\ref{fig:stark9} shows the energy dependence of the Stark cross-sections
for $n=9$. The classical-trajectory model and fixed field model are in agreement 
with each other 
for kinetic  energies above 10~eV (molecular target) and 2~eV (atomic target). At lower
energies where the Coulomb transitions make up a substantial part of the cross-sections,  the
fixed field model overestimates the Stark cross-sections.
%%%%%%%%%%%%%%%%%%%%%%%%%%%%%%%%%%%%%%%%%%%%%%%%%%%%%%%%%%%%%%%%%%%%%%%%%
%%%%%%%%%%%%%%%%%%%%%%%%%%%%%%%%%%%%%%%%%%%%%%%%%%%%%%%%%%%%%%%%%%%%%%%%%

\begin{figure}
\center{
\epsfig{file=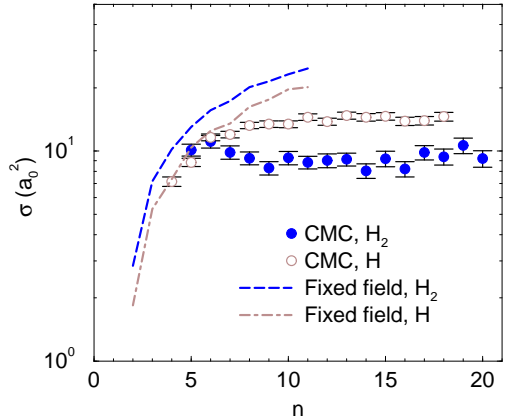, width=6.5cm}
}
\caption{
The $n$ dependence of the transport cross-sections for muonic hydrogen 
at $T=1$~eV.  
 The results of the classical Monte
Carlo model with molecular (filled circles) and atomic target (circles) are shown in comparison with the 
semiclassical fixed field model.
}
\label{fig:cmup3}
\end{figure}

The Stark mixing and elastic scattering processes, 
(\ref{EqStMol}) and (\ref{EqStAt}), 
lead to a deceleration of the exotic atom. Their
importance in the kinetics of atomic cascade 
can be estimated with  the  corresponding transport cross-section
\be
  \sigma_n^\mathrm{tr}=\int (1-\cos \theta)\frac{\dd\sigma_{n\to n}}{\dd\Omega}\dd\Omega
\ee
where $\mathrm{d}\sigma_{n\to n}/\mathrm{d}\Omega$ is the differential cross-section for 
the processes~(\ref{EqStMol}) or (\ref{EqStAt}) averaged over $l$. This estimate based
on  the  transport cross-section 
neglects the Coulomb deexcitation process which can lead to both deceleration and 
acceleration, and,  in the case of molecular target, the additional deceleration
due to excitation of the $\Ha_2$ molecule.
The $n$ dependence of the transport cross-sections at 1~eV for muonic hydrogen scattering
from hydrogen atoms and molecules is shown in Figure~\ref{fig:cmup3}. There is a fair 
agreement 
between the CMC and the fixed field model for atomic target below $n\sim 8$. For higher
$n$,  the inelastic effects due to  the Coulomb deexcitation process become important,
 and the
fixed field model overestimates the transport cross-section. For molecular target, the
discrepancy between the two models is larger because the Coulomb cross-section
makes up a larger fraction of the total cross-section as compared to the CMC model with
atomic target (for $n=10$ and $T=1$~eV the fractions are~$\sim 0.24$ for molecular target
and $\sim 0.11$ for atomic target). 
%The fixed field model and other models~\cite{jensen02epjd} 
% based on expansion of the internal $x^-p$ wave function into the set of eigenfunctions
% with the same $n$ do, therefore, not give a realistic description of the slowing down
% except for the final part of the cascade ($n<6$).

\begin{figure}
\center{
\epsfig{file=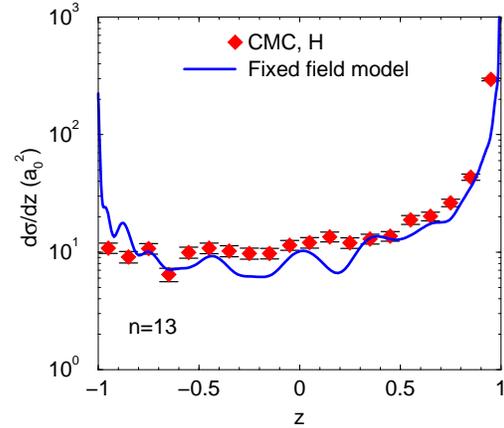, width=6.5cm}
}
\caption[Differential cross-sections for \mup]{
Differential  cross-section $\dd\sigma/\dd z$ for muonic hydrogen
for $n_i=13$ and laboratory kinetic  energy $T=1$~eV.   
The classical Monte Carlo results for atomic target % (6000 trajectories)
are shown with filled diamonds and the curve corresponds to the semiclassical 
fixed field model for atomic target.
}
\label{fig:ztotal}
\end{figure}

%  We found a good agreement between the two classical Monte
%Carlo models and their corresponding semiclassical counterparts 
%for the $l$-averaged differential cross-sections.  
Figure~\ref{fig:ztotal} shows the $l$-averaged differential cross-section
(using 20 equally spaced bins in $z$) 
 summed over all the final channels for $n_i=13$ 
in the classical Monte Carlo model with atomic target.  
The cross-section  is in  good agreement with that of the semiclassical 
fixed field model.  The pattern of maxima and minima in the 
semiclassical differential cross-sections is a characteristic feature of 
quantum mechanical scattering, which, of course,  cannot be reproduced in a 
classical model.

%%%%%%%%%%%%%%%%%%%%%%%%%%%%%%%%%%%%%%%%%%%%%%%%%%%%%%%%%%%%%%%%%%%%%%%%%%%%%
%%%%%%%%%%%%%%%%%%%%%%%%%%%%%%%%%%%%%%%%%%%%%%%%%%%%%%%%%%%%%%%%%%%%%%%%%%%%%
%%%%%%%%%%%%%%%%%%%%%%%%%%%%%%%%%%%%%%%%%%%%%%%%%%%%%%%%%%%%%%%%%%%%%%%%%
%%%%%%%%%%%%%%%%%%%%%%%%%%%%%%%%%%%%%%%%%%%%%%%%%%%%%%%%%%%%%%%%%%%%%%%%%
\begin{figure}
\center{
\epsfig{file=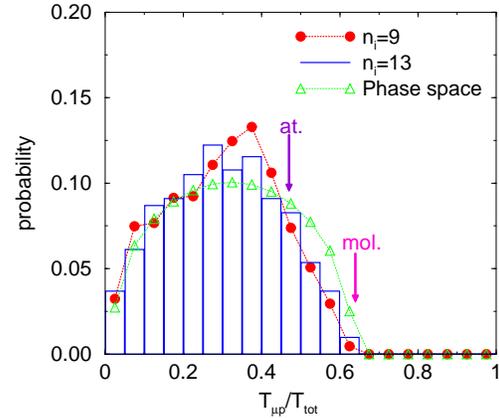, width=6.5cm}
}
\caption{
Distribution over the \mup\ kinetic energy as a fraction of the total
energy in the final state, $T_{\mu^- p}/T_\mathrm{tot}$, for the Coulomb deexcitation of 
muonic hydrogen in the initial states $n_i=9,13$ 
at the laboratory kinetic energy $T=1$~eV. % (10000 trajectories for each case). 
The phase space distribution is shown for comparison. The vertical arrows
indicate   the \mup\ final energies  of the two-body final states 
$\mup+\Ha$ (at.) and $\mup+\Ha_2$ (mol.). 
}
\label{fig:Tmup}
\end{figure}

The kinetic energy of the $x^-p$ in the final state is important for
detailed cascade calculations.     
Let $T_{x^-p}$, $T_\mathrm{H}$ and $T_{\mathrm{H}_2}$ be the
CMS kinetic energies of the $x^-p$, the H (for atomic target),
and the $\mathrm{H}_2$ (for molecular target).
The total kinetic energy is shared among the two ($x^-p$ and H) 
or three atoms ($x^-p$ and two hydrogen atoms): 
\be
 T_\mathrm{tot}=\left\{
\begin{array}{ll}
  T_{x^-p}+T_\mathrm{H},&{\rm atomic\; target}\\
  T_{x^-p}+T_{\mathrm{H}_2}+\Delta E_\mathrm{target},&{\rm molecular\; target }
\end{array}\right.
\ee
In the case of atomic target, the energy of the $x^-p$ in CMS is fixed:
\be
 \frac{T_{x^-p}}{T_\mathrm{tot}}=\frac{M_\mathrm{H}}{M_{xp}+M_\mathrm{H}}\quad(=0.47\;\mathrm{for}\;\mup )
\label{eq:txp_ttot}
\ee 
where $M_\mathrm{H}$ and $M_{xp}$ are the masses of the hydrogen atom and the \xp\ atom,
correspondingly. The case of   molecular target corresponds to a three-body final state 
with the kinematical boundaries: 
\be
  0\leq \frac{T_{x^-p}}{T_\mathrm{tot}}\leq
  \frac{T_{x^-p}^\mathrm{max}}{T_\mathrm{tot}}
  =\frac{2M_\mathrm{H}}{M_{xp}+2M_\mathrm{H}}.
\ee
The upper boundary (0.64 for muonic hydrogen) is reached when
the hydrogen molecule remains in its ground state corresponding effectively to
a two-body ($(\xp)+(\Ha_2)$) final state. 
Figure~\ref{fig:Tmup} shows the distributions in $T_{x^-p}/T_\mathrm{tot}$ 
for Coulomb deexcitations calculated in the classical Monte Carlo model
for muonic hydrogen with $n_i=9$, 13 and $T=1$~eV. 
The approximation of  effective two-body final states clearly fails, whereas  
 the pure phase space distribution 
\be
  f( T_{x^-p}) = 
   \frac{4T_\mathrm{tot}}{\pi T_{x^-p}^\mathrm{max}}\sqrt{1-
   \left(\frac{2T_{x^-p}}
   {T_{x^-p}^\mathrm{max}}-1\right)^2}
\label{T_phasespace}
\ee
gives a fair description of the results. 
%Further studies show 
%that the CMS distribution in the final kinetic energy of the 
%exotic atom can be fairly approximated by Eq.~(\ref{T_phasespace})
%for a broad range of the initial quantum numbers $n$ and kinetic 
%energies $T$.

%%%%%%%%%%%%%%%%%%%%%%%%%%%%%%%%%%%%%%%%%%%%%%%%%%%%%%%%%%%%%%%%%%%%%%%%%%%
\subsection{Antiprotonic hydrogen}
\label{sect:res:pbp}

\begin{figure}[htb]
\center{
\epsfig{file=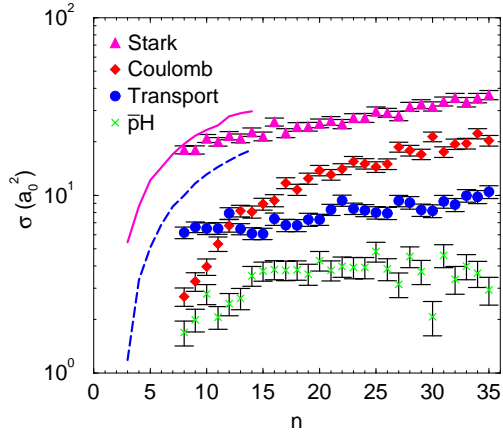, width=6.5cm}
}
\caption[Cross-sections for \pbp]{
The $n$ dependence  of the cross-sections of the cascade processes in 
antiprotonic hydrogen calculated in 
the classical-trajectory Monte Carlo model for molecular target.
The laboratory energy is $T=1$~eV.  
The Stark mixing and transport cross-sections calculated in the fixed field model
for molecular target
are shown with solid and dashed lines, respectively.
}
\label{fig:pbpcn:n}
\end{figure}

\begin{figure}[htb]
\center{
\epsfig{file=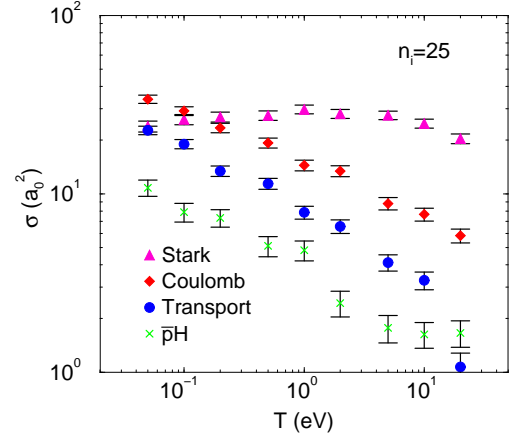, width=6.5cm}
}
\caption[Cross-sections for \pbp]{
The energy dependence of the cross-sections of the cascade processes in 
antiprotonic hydrogen in the state $n_i=25$ calculated in 
the classical-trajectory Monte Carlo model for molecular target.
}
\label{fig:pbpcn:t}
\end{figure}

The atomic cascade in antiprotonic hydrogen starts around $n_i\sim 30$; 
thus classical mechanics is even a better approximation 
than in the muonic hydrogen case.  
The $n$ dependence of the Stark mixing,  Coulomb deexcitation, transport,
and the $\bar{p}\Ha$ formation cross-sections 
is shown in Figure~\ref{fig:pbpcn:n}, and the  energy dependence is 
demonstrated in Figure~\ref{fig:pbpcn:t}. As with muonic hydrogen, the fixed field
model overestimates the Stark mixing and especially the transport cross-section
because the inelasticity effects due to Coulomb deexcitation are 
 not included in this framework.
%%%%%%%%%%%%%%%%%%%%%%%%%%%%%%%%%%%%%%%%%%%%%%%%%%%%%%%%%%%%%%%%%%%%%%%%%%%%%
%%%%%%%%%%%%%%%%%%%%%%%%%%%%%%%%%%%%%%%%%%%%%%%%%%%%%%%%%%%%%%%%%%%%%%%%%
%%%%%%%%%%%%%%%%%%%%%%%%%%%%%%%%%%%%%%%%%%%%%%%%%%%%%%%%%%%%%%%%%%%%%%%%%%%%%
%%%%%%%%%%%%%%%%%%%%%%%%%%%%%%%%%%%%%%%%%%%%%%%%%%%%%%%%%%%%%%%%%%%%%%%%%

\begin{figure}[htb]
\center{
\epsfig{file=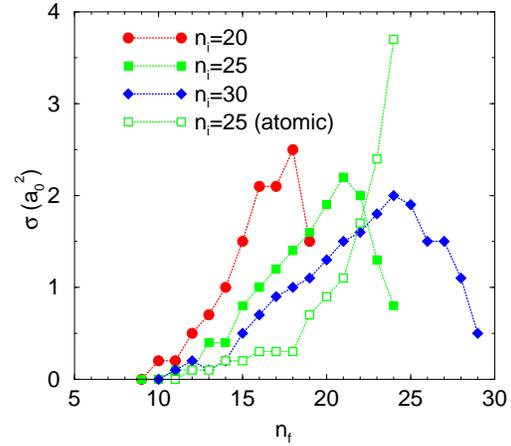, width=6.5cm}
}
\caption{
 The  $n_f$ dependence of the Coulomb cross-sections 
 for antiprotonic hydrogen for $n_i=20,25,30$ and the laboratory kinetic 
 energy $T=1$~eV.
For the sake of clarity we do not show the statistical error bars.
}
\label{fig:dnpp}
\end{figure}

  Figure~\ref{fig:dnpp} shows the distribution over the final states $n_f$ 
for the Coulomb deexcitation of the antiprotonic hydrogen at 
the laboratory  energy $T=1$~eV.  For high $n$ initial states, the most probable 
Coulomb transitions are the ones with a large change of the principal 
quantum number ($\Delta n\gg 1$), with the molecular target being 
essential for this feature.  A very important consequence of this result is 
that at the beginning of the atomic cascade a small number of Coulomb transitions 
is sufficient to bring the \pbp\ to the middle stage, where, depending 
on the target density, the radiative or Auger deexcitation takes over.  

%The  distribution over final $n_f$ for the Coulomb cross-sections 
%with molecular target is determined by the interatomic H-H
%potential, Eq.~(\ref{vmorse}). By varying the potential depth, $D_e$, in the Morse
%potential one finds that a weaker (stronger) potential shifts the maximum of the distribution
%towards smaller (larger) values of $\Delta n$. Figure~\ref{fig:dnpp}b shows the distribution 
%over final $n_f$ for  $n_i=25$ at $T=1$~eV and $D_e=0.1$~eV, 10~eV, 100~eV.  
%A collision with a  hydrogen molecule is, therefore, qualitatively different 
%from two independent collisions with hydrogen atoms.  
%%%%%%%%%%%%%%%%%%%%%%%%%%%%%%%%%%%%%%%%%%%%%%%%%%%%%%%%%%%%%%%%%%%%%%%%%%%%%
%%%%%%%%%%%%%%%%%%%%%%%%%%%%%%%%%%%%%%%%%%%%%%%%%%%%%%%%%%%%%%%%%%%%%%%%%
\begin{figure}[htb]
\center{
\epsfig{file=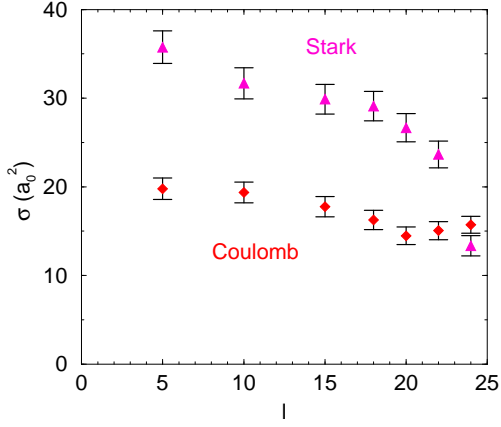, width=6.5cm}
}
\caption{
The $l_i$ dependence of the Coulomb (filled diamonds)  and Stark (filled triangles)
cross-sections
 for antiprotonic hydrogen for $n_i=25$ and  $T=1$~eV. The results are calculated
  in the classical  Monte Carlo model with molecular target.
}
\label{fig:lpp}
\end{figure}

%
% In contrast to this picture, the conventional cascade model required ... 
%
  The dependence of the Coulomb cross-sections on
the angular momentum $l_i$ of the initial state is weak, 
see Figure~\ref{fig:lpp} for antiprotonic hydrogen with $n_i=25$.
The Stark cross-sections show a moderate dependence on $l_i$: 
they are smaller for the circular states ($l_i=n_i-1$) than for 
the low $l_i$, by about 50\%.  The reason for this is  that the elongated
ellipses  in the low $l$ states are more easily perturbed by the electric field 
of the target molecule. A similar effect is expected if a quantum mechanical
description of the  \pbp\ is used: the size of the \pbp\ as estimated by
the expectation value of $r^2$ is given by
\be
 \langle r^2 \rangle=\frac{n^2}{2\mu_{\pbp}^2}\left( 5n^2+1-3l(l+1) \right).
\ee
For high $n$ states, the expectation value of $r^2$ for the circular state is only
40\% of that of the $ns$ states.

%%%%%%%%%%%%%%%%%%%%%%%%%%%%%%%%%%%%%%%%%%%%%%%%%%%%%%%%%%%%%%%%%%%%%%%%%%%
\section{External Auger effect in the eikonal approximation}
\label{sect:auger}

   In our treatment of the Coulomb and Stark mixing collisions in 
Section~\ref{sect:ctmc}, the electronic degrees of 
freedom were assumed to be frozen.  These degrees of freedom, however, 
play an important role in the Auger deexcitation process
\be
 (x^-p)_{n_il_i}+\mathrm{H}\to(x^-p)_{n_fl_f}+p+e^-
 .  \label{auger}
\ee 
The Auger transitions are often treated in the Born approximation~\cite{leon62} 
that gives (conveniently) energy independent rates.  However, this approximation 
violates unitarity for some important ranges of principal quantum numbers and 
kinetic energies.    
For kinetic energies in the range of few eV, the eikonal 
approximation~\cite{bukhvostov82} provides a more suitable framework.  
In this section, we use the eikonal approach to calculate Stark mixing and 
Auger deexcitation simultaneously.  As a result, the corresponding partial 
wave cross-sections are  consistent with unitarity.  

The cross-section for the process~(\ref{auger}) was calculated in~\cite{bukhvostov82} 
by assuming that the exotic atom moves along a straight line trajectory 
with constant  velocity $v$ through the electric field of the hydrogen atom at rest.  
The cross-section is given by
\be
 \sigma^\mathrm{Auger}_{n_il_i}=2\pi\int_0^\infty P(\rho)\rho \dd\rho
 \label{augerbp1}
\ee
where $P(\rho)$ is the reaction probability for the impact parameter $\rho$: 
\be
P(\rho)=1-e^{-I(\rho)},\quad 
I(\rho)=\frac{1}{v}\int_{-\infty}^{\infty}\Gamma_{n_il_i}(\sqrt{\rho^2+z^2})\dd z.
\label{augerbp2}
\ee
The reaction rate, $\Gamma_{n_il_i}(R)$,  at distance $R$ 
 is the sum of the partial  rates  $\Gamma_{n_il_i\to n_fl_f}(R)$ over all final states.  
According to \cite{bukhvostov82} the estimated rates are 
\be
 \Gamma_{n_il_i\to n_fl_f}(R)=\gamma\frac{1}{(R^2+b^2)^3}+\gamma_1\frac{k_e^2}{1+k_e^2}
\exp(-2R)
\ee  
where $k_e$ is the electron momentum, $b=1.5$, and the parameters 
$\gamma$ and $\gamma_1$ are given by  
\begin{eqnarray}
 \gamma & = & 
   \frac{2^{10}\pi}{3}\mu_{xp}^{-2}\frac{\exp\left((-4/k_e)\arctan k_e\right)}
   {(1+k_e^2)^6(1-\exp(-2\pi/k_e))}\nonumber\\
   &\times&(C^{l_f0}_{l_i010})^2(R_{n_il_i}^{n_fl_f})^2,
\\
 \gamma_1 & = & 
   \frac{16}{3k_e}\muxp^{-2}(C^{l_f0}_{l_i010})^2(R_{n_il_i}^{n_fl_f})^2
\end{eqnarray}
where $C^{l_f0}_{l_i010}$ is a Clebsch-Gordan coefficient and $R_{n_il_i}^{n_fl_f}$ is the
radial matrix element~\cite{bethe57}. 
The transition rate is proportional to the square of the dipole matrix element,  
therefore only transitions with $\Delta l=|l_f-l_i|=1$ are possible. 
    
  The Auger deexcitation rate, as a function of $n$, peaks at the $n$-value where
the energy released in a $\Delta n=1$ transition is just sufficient to ionize
the hydrogen atom.  
  The effect of these high-rate Auger transitions is that
   the inelastic cross-sections  for some partial 
waves are not small in comparison with the 
unitarity limit. Therefore the corresponding inelasticity should be taken 
into account in the calculations of other collisional processes.  
One can expect that taking the Auger effect into account will reduce 
the other inelastic cross-sections.       
   In order to examine this effect, we include the Auger deexcitation 
in the framework presented in~\cite{jensen02epjd} for  calculating  
Stark mixing and elastic scattering.
In the same way as the nuclear absorption processes in hadronic atoms 
were taken into  account via imaginary energy shifts of the $ns$-states,
the Auger deexcitation process is included via the imaginary 
absorption potential, $-i\Gamma_{nl}(R)/2$.
%We use the fixed field approximation where the electric field of the target 
%atom is assumed to be directed along the quantization axis of the exotic
%atom.  
%The method employed follows chapter~\ref{ch:lown} 
%where the nuclear absorption processes in hadronic atoms were taken into 
%account via imaginary energy shifts of the $ns$-states.  
%In the same way, the Auger deexcitation process is included via the imaginary 
%absorption potential, $-i\Gamma_{nl}(R)/2$.  
The calculations can be done in the close-coupling model, the 
semiclassical model, and the fixed field model. 
In the case of the fixed field model, 
the time-dependent Schr{\"o}dinger equation for the set of 
the linear independent solutions forming the $n^2\times n^2$ matrix $A$ is given by    
\be
  i\dot{A}(t)=H(t)A(t)
 \label{adot}
\ee
where the interaction is given by
\begin{eqnarray}
 H(t)&=&Z \frac{1}{R^2(t)}(1+2R(t)+2R^2(t))e^{-2R(t)}\nonumber\\
 &+&\Delta E
 -i\frac{\Gamma(R)}{2}.
\label{hamilton}
\end{eqnarray}
Here $\Delta E$ is a diagonal matrix corresponding to the energy shifts due 
to the vacuum polarization and the strong interaction.
The term $\Gamma(R)$ is a diagonal matrix with the matrix elements $\Gamma_{nl}(R)$. 
The factor $Z$ originates from the dipole operator and has the following 
matrix elements ($i=|nl\Lambda\rangle$, $j=|n(l-1)\Lambda\rangle$): 
\be
  Z_{ij}=
      -\frac{3n}{2\mu_{xp}}\sqrt{\frac{(l^2-\Lambda^2)(n^2-l^2)}{(2l+1)(2l-1)}}.
\ee

The solution of Equation (\ref{adot}) using the method described in  \cite{jensen02epjd}
gives the scattering matrix $S^\mathrm{FF}$. 
The cross-sections for the transitions $n_il_i\to n_il_f$ are given by
\begin{eqnarray}
 \sigma_{n_il_i\to n_il_f}&=&\frac{1}{2l_i+1}\frac{\pi}{k^2}\sum_{J}(2J+1)\nonumber\\
 \times&\sum_{\Lambda}&
|\langle n_i;JM \Lambda l_f|S^\mathrm{FF}-1|n_i;JM\Lambda l_i\rangle|^2
\end{eqnarray}
and the ones of the Auger deexcitation by
\begin{eqnarray}
 \sigma_{n_il_i}^\mathrm{Auger}&=&\frac{1}{2l_i+1}\frac{\pi}{k^2}\sum_{J}(2J+1)
\Big( (2l_i+1)\nonumber\\
&-&\sum_{\Lambda l_f}|\langle n_i;JM \Lambda l_f|S^\mathrm{FF}
|n_i;JM\Lambda l_i\rangle|^2\Big).
\label{sigma_auger}
\end{eqnarray}
We will refer to this framework as the eikonal multichannel model.

%%%%%%%%%%%%%%%%%%%%%%%%%%%%%%%%%%%%%%%%%%%%%%%%%%%%%%%%%%%%%%%%%%%%%%%%%%%%%
\begin{figure}
\mbox{
\epsfig{file=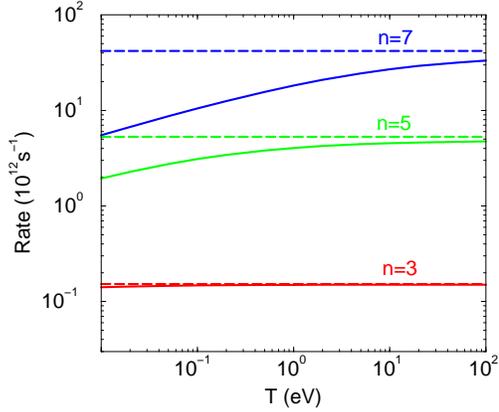, width=6.5cm}
}
\caption{
 The energy dependence of the Auger deexcitation rates for muonic hydrogen 
in liquid hydrogen. The results of the eikonal approximation
are shown with solid lines and those of the Born approximation
with dashed lines. 
}
\label{fig:auger:t}
\end{figure}
%%%%%%%%%%%%%%%%%%%%%%%%%%%%%%%%%%%%%%%%%%%%%%%%%%%%%%%%%%%%%%%%%%%%%%%%%
%%%%%%%%%%%%%%%%%%%%%%%%%%%%%%%%%%%%%%%%%%%%%%%%%%%%%%%%%%%%%%%%%%%%%%%%%%%%%
\begin{figure}
\mbox{
\epsfig{file=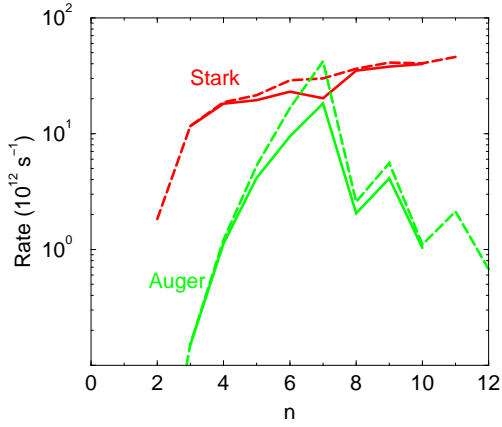, width=6.5cm}
}
\caption{
The $n$ dependence of the Auger deexcitation and Stark mixing rates at 1~eV 
for muonic hydrogen in liquid hydrogen.  
The results of the eikonal multichannel model 
are shown with solid lines.  The Auger deexcitation rates calculated 
in the  the Born approximation and the Stark mixing rates obtained 
without Auger deexcitation are shown with dashed lines.  
}
\label{fig:auger:n}
\end{figure}
%%%%%%%%%%%%%%%%%%%%%%%%%%%%%%%%%%%%%%%%%%%%%%%%%%%%%%%%%%%%%%%%%%%%%%%%%

%%%%%%%%%%%%%%%%%%%%%%%%%%%%%%%%%%%%%%%%%%%%%%%%%%%%%%%%%%%%%%%%%%%%%%%%%%%
\begin{figure}
\mbox{
\epsfig{file=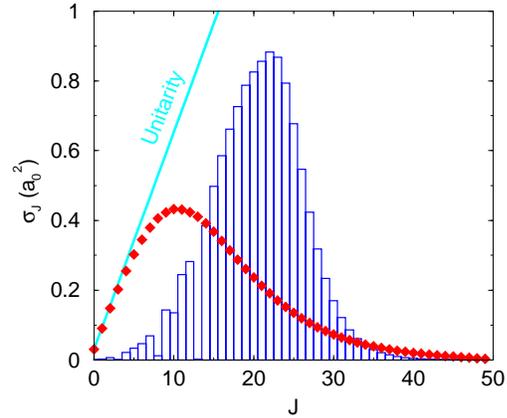, width=6.5cm}
}
\caption{
 The $J$ dependence of the $l$-average partial wave cross-sections for
muonic hydrogen for $n=7$ and laboratory kinetic energy $T=3$~eV.  
The cross-sections for Auger deexcitation with $\Delta n=1$ are shown 
with diamonds, those of Stark mixing with histograms, 
and the unitarity limit with a thick solid line. 
}
\label{fig:auger:pw}
\end{figure}
%%%%%%%%%%%%%%%%%%%%%%%%%%%%%%%%%%%%%%%%%%%%%%%%%%%%%%%%%%%%%%%%%%%%%%%%%%%
%%%%%%%%%%%%%%%%%%%%%%%%%%%%%%%%%%%%%%%%%%%%%%%%%%%%%%%%%%%%%%%%%%%%%%%%%%%
\begin{figure}
\mbox{
\epsfig{file=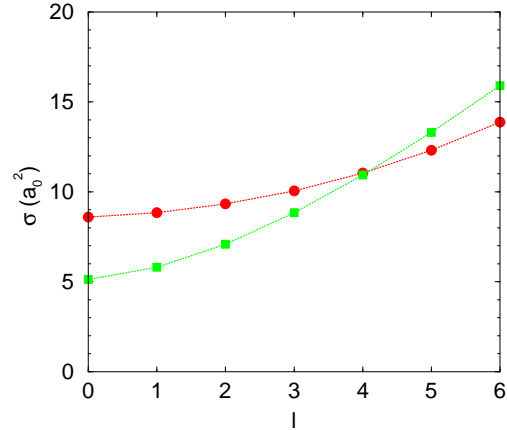, width=6.5cm}
}
\caption{
  The initial $l$ dependence of the Auger deexcitation cross-sections at 1~eV 
for muonic hydrogen at $n=7$. The result of the  eikonal multichannel model 
is shown with filled circles and that of  method \cite{bukhvostov82} with filled squares.
}
\label{fig:auger:l7}
\end{figure}
%%%%%%%%%%%%%%%%%%%%%%%%%%%%%%%%%%%%%%%%%%%%%%%%%%%%%%%%%%%%%%%%%%%%%%%%%%%

The $l$-average Auger deexcitation cross-sections calculated with the method 
of~\cite{bukhvostov82} (Equations (\ref{augerbp1}) and (\ref{augerbp2})) 
agree closely with our results in the eikonal multichannel model. 
Figure~\ref{fig:auger:t} shows the $l$-average $\Delta n=1$ Auger deexcitation 
rates in  muonic hydrogen for $n=3,5,7$. The rates have been calculated 
in the eikonal approximation and the Born approximation.   
  The rates in the eikonal approximation are lower in the low energy range, 
but they approach the ones of the Born approximation for high energies.  
The $n$ dependence of the Auger deexcitation and Stark mixing rates for 
muonic hydrogen is presented in Figure~\ref{fig:auger:n}.  
The two approaches are in a fair agreement with each other except for  
the states $n=6,7$ where the Auger rates have the highest values.  
For the state $n=7$, the Stark mixing rates are reduced by almost 50\% 
when the inelasticity due to the Auger effect is included.
This resembles the situation with the eikonal and the Born approximations which 
disagree when the Auger deexcitation cross-sections are large, 
in which case the eikonal approximation gives smaller cross-sections than 
the Born approximation. 
The explanation of this effect is given in Figure~\ref{fig:auger:pw}
showing the average partial wave cross-sections for the collision $(\mu p)_{n=7}+\Ha$. 
The Auger deexcitation cross-sections are saturated in 
the low angular momentum region and, therefore, the Born approximation fails.  
  Though the $l$-average results agree for the two eikonal approaches,  
the $l$ dependence of the cross-sections in the eikonal multichannel model 
is weaker because of the effect of Stark mixing as demonstrated  
in Figure~\ref{fig:auger:l7}.

The eikonal  approximation as described above 
does not give the differential cross-section and
distribution over final $l_f$   for the Auger transitions. The partial wave 
cross-sections, Figure~\ref{fig:auger:pw}, show that the main contribution to
the Auger cross-section comes from low partial waves, {\it i.e.} from the strong
mixing region. This suggests that the distribution in  $l_f$ is nearly statistical and 
that the differential cross-section is less forward-peaked than the 
elastic and Stark differential cross-sections~\cite{jensen02epjd}.

%%%%%%%%%%%%%%%%%%%%%%%%%%%%%%%%%%%%%%%%%%%%%%%%%%%%%%%%%%%%%%%%%%%%%%%%%%%
\section{Conclusions}
\label{sect:conc}

The collisional deexcitation mechanisms of the exotic hydrogen atoms 
in highly excited states have been investigated in detail using the 
classical-trajectory Monte Carlo method.  
The Coulomb transitions have been shown to be the dominant mechanism of 
collisional deexcitation of highly excited exotic atoms.  
Target molecular structure has large effects on the Coulomb deexcitation.  
In particular, the distribution over the final states favors large change 
of the principal quantum number $n$ contrary to the case of
atomic target.  
This feature is very important for the cascade kinetics as it leads to 
a fast deexcitation and a significant acceleration at the initial stage 
of the atomic cascade \cite{jensen02next}.  
The calculated cross-sections provide a more reliable theoretical input for 
further cascade studies by removing the long standing puzzle of the so-called 
chemical deexcitation \cite{leon62}, which was used, on  purely phenomenological 
grounds, in many cascade calculations without clarification of 
underlying dynamics.   

The external Auger effect has been studied in an eikonal multichannel
model which allows us to calculate Stark mixing, elastic scattering,
and Auger deexcitation simultaneously. Partial wave cross-sections computed 
in this framework are consistent with unitarity. For  ranges of principal quantum 
numbers and kinetic energies where the unitarity constraint is important, 
the Auger cross-sections computed
in this model are significantly lower than those of the Born approximation~\cite{leon62}.

The first results of  cascade calculations using the cross-sections of~\cite{jensen02epjd} 
and the present paper 
have been presented in~\cite{markushin02hyp,jensen02pin}.
More detailed results of the cascade calculations
will be discussed in a separate publication~\cite{jensen02next}.      

\section*{Acknowledgment}

We thank F.~Kottmann, L.~Simons, D.~Taqqu, and R.~Pohl for 
fruitful and stimulating discussions.

%%%%%%%%%%%%%%%%%%%%%%%%%%%%%%%%%%%%%%%%%%%%%%%%%%%%%%%%%%%%%%%%%%%%%%%%%%%
\newcommand{\etal}{\mbox{\it et al.}}

\bibliographystyle{unsrt}

\end{document}